\documentclass[PHM, 2026]{PHMSociety}

\usepackage{graphicx}
\usepackage{amsmath}
\usepackage{amssymb}
\usepackage{url}
\usepackage{booktabs}
\usepackage{multirow}
\usepackage{array}
\usepackage{xcolor}
\usepackage{subcaption}
\usepackage[acronym,nomain,nonumberlist,nopostdot]{glossaries}
\glsdisablehyper
\usepackage[capitalize,noabbrev,nameinlink]{cleveref}

\crefname{equation}{Eq.}{Eqs.}
\Crefname{equation}{Eq.}{Eqs.}

\newacronym{phm}{PHM}{prognostics and health management}
\newacronym{sic}{SiC}{silicon carbide}
\newacronym{mosfet}{MOSFET}{metal-oxide-semiconductor field-effect transistor}
\newacronym{cm}{CM}{condition monitoring}
\newacronym{eol}{EOL}{end-of-life}
\newacronym{rul}{RUL}{remaining useful life}
\newacronym{aqg}{AQG}{Automotive Qualification Guideline}
\newacronym{ecpe}{ECPE}{European Center for Power Electronics}
\newacronym{dut}{DUT}{device under test}
\newacronym{pof}{PoF}{physics of failure}
\newacronym{mlp}{MLP}{MultiLayer Perceptron}
\newacronym{cmlp}{CMLP}{Contextual MLP}
\newacronym{node}{NODE}{Neural ODE}
\newacronym{mae}{MAE}{mean absolute error}
\newacronym{nl}{NL}{Norris--Landzberg}
\newacronym{cv}{CV}{coefficient of variation}
\newacronym{piml}{PIML}{physics-informed machine learning}
\newacronym{cnn}{CNN}{convolutional neural network}
\newacronym{lstm}{LSTM}{long short-term memory network}

\begin{document}

% Paper Title
\title{Physics-Informed Condition Monitoring of {SiC} Power Modules}

% Authors List
\author{%
    Mattia Scarpa\authorNumber{1,2,4},
    Evgeny Kusmenko\authorNumber{2},
    Francesco Toso\authorNumber{4},
    Mattia Bruschetta\authorNumber{1},
    Ruggero Carli\authorNumber{1}, 
    and Simon Achatz\authorNumber{3}
}

% Author Affiliations
\address{%
    \affiliation{1}{%
        Department of Information Engineering,
        University of Padova,
        Padova, Italy}{%
        {\email{scarpamatt@dei.unipd.it}}\\
        {\email{carlirug@dei.unipd.it}}\\
        {\email{bruschet@dei.unipd.it}}
    }
    \tabularnewline
    \affiliation{2}{%
        Infineon Technologies Dresden GmbH \& Co.\ KG,
        Dresden, Germany}{%
        {\email{Evgeny.Kusmenko@infineon.com}}
    }
    \tabularnewline
    \affiliation{3}{%
        Infineon Technologies AG,
        Munich, Germany}{%
        {\email{Simon.Achatz@infineon.com}}
    }
    \tabularnewline
    \affiliation{4}{%
        Newtwen S.r.l.,
        Padova, Italy}{%
        {\email{francesco.toso@newtwen.com}}
    }
}

% Create the title
\maketitle

% PHM Society Distribution License
\phmLicenseFootnote{Mattia Scarpa}

% ---- arXiv preprint notice -------------------------------------------
\renewcommand{\thefootnote}{}%
\footnotetext{Preprint. This work has been accepted for presentation at the
Annual Conference of the Prognostics and Health Management Society 2026
(PHM~2026). This document is the authors' extended version of the accepted
paper and includes additional material not present in the conference version.}%
\renewcommand{\thefootnote}{\arabic{footnote}}\setcounter{footnote}{0}%
% ----------------------------------------------------------------------

% Abstract
% Note: do not cite references in the abstract per PHM Society guidelines.
\begin{abstract}
Silicon carbide (SiC) power modules are increasingly deployed in automotive
traction inverters, where reliable condition monitoring is essential to prevent
in-service failures and reduce maintenance costs.
Despite extensive qualification procedures standardized under AQG 324, no
consolidated approach exists for in-field health state estimation of SiC
devices.
Existing methods range from physics-of-failure lifetime models, which lack
real-time applicability, to purely data-driven architectures that require large
labeled datasets and offer limited generalization, to physics-informed machine
learning frameworks that, while promising, remain computationally demanding for
embedded deployment.

This work addresses condition monitoring of SiC MOSFET power modules assembled
with sintered packaging technology, which prevents solder degradation and thereby
produces aging behavior qualitatively distinct from previously studied devices.
In solder-based modules, the forward voltage drop $V_{DS}$ follows smooth
quasi-exponential trajectories driven by progressive solder delamination.
With sintered packaging this mechanism is suppressed, and $V_{DS}$ instead
exhibits multi-regime degradation profiles; modules are additionally subject to
wirebond liftoff events that introduce abrupt, non-monotonic perturbations
directly onto $V_{DS}$, posing new challenges for  condition
monitoring approaches.

To address these challenges, we propose a condition monitoring framework
integrating three complementary elements.
First, physics-informed feature engineering replaces raw sensor signals with
physically grounded inputs, including cumulative damage indicators derived from
junction temperature swing and mean junction temperature, combined with a Miner
rule accumulator, which encode degradation history in a form directly
interpretable by the model.
Second, a monotonicity constraint enforced via gradient penalty regularization
embeds the expected degradation direction as a physics-guided prior, improving
generalization and cross-validation stability.
Third, the model output is parameterized as a heavy-tailed distribution rather
than a point estimate, providing calibrated uncertainty quantification
inherently robust to the out-of-distribution variance introduced by liftoff
events.

The framework is evaluated on an industrial power cycling dataset provided by
Infineon Technologies, acquired under multiple operating conditions.
Multiple neural network architectures of varying complexity and sequential
context are compared within a rigorous cross-validation protocol.
The full physics-informed feature set combined with gradient penalty
regularization consistently outperforms purely data-driven baselines, reducing
mean absolute error by approximately 70\% and maintaining stable performance
across all cross-validation folds, including conditions where baseline models
degrade significantly.
These results confirm that integrating physical prior knowledge at the feature,
constraint, and output distribution levels is both necessary and sufficient to
handle the complexity of SiC power module degradation, while remaining
lightweight enough for practical embedded deployment.
\end{abstract}

% ============================================================
\section{Introduction}
\label{sec:intro}
% ============================================================
\Gls{sic} power \glspl{mosfet} are now widely deployed in the
high-voltage stage of automotive traction inverters, where their higher
switching frequency and elevated junction-temperature limits allow for
smaller passive components and higher converter efficiency. Reliability
of the power-semiconductor stage has long been identified by industry
as one of the most pressing reliability concerns among the components
that make up a power converter~\shortcite{yang2011}, and accelerated
power-cycling tests under the \gls{ecpe} \gls{aqg}~324
qualification~\shortcite{aqg324} provide the bridge between bench
characterization and field deployment. The in-service \gls{cm} problem
is distinct from the qualification one: rather than counting cycles to
a fixed threshold under a known stress profile, on-board electronics
must keep an updated health estimate from electrical measurements taken
while the module is operating under variable load. Reviews of \gls{sic}
reliability converge on the dominant failure modes, but the electrical
signals proposed for tracking those modes online are still treated as
candidates rather than as a consolidated indicator set, and their
combination into a per-cycle health estimate consistent with the
lifetime models obtained from accelerated tests is identified as a
central open methodological problem~\shortcite{ni2020}.

The chip-solder degradation mechanism has long been recognized as
the leading source of package-level failure in conventional power
modules under thermo-mechanical
cycling~\shortcite{clech1997,kovacevic2010,kovacevic2015}, with a
physical signature on the on-state voltage drop that is well
described as a smooth quasi-exponential rise driven by progressive
crack growth in the die-attach. This profile contrasts sharply with
the abrupt step-like behavior of wirebond fatigue, where each
individual bond-wire lift-off has been observed to produce a
stepwise increase in the on-state voltage
drop~\shortcite{kumar2026}. For this reason, data-driven \gls{sic}
\gls{cm} has predominantly grown on modules where solder
degradation is the dominant failure mode, and has gradually
drifted away from the closed-form parametric route that fits an
electro-thermal indicator to a regression law and inverts it to
recover a lifetime fraction~\shortcite{diNuzzo2022,diNuzzo2023}. The
analytical transparency of that route comes at the cost of a tight
coupling to the assumed indicator shape, and more flexible
machine-learning formulations have also been explored in parallel,
including sequence models that forecast the relative on-state
voltage drop forward to the qualification
threshold~\shortcite{olschewski2025,villalobos2025} and end-to-end
snapshot models that map the current measurement directly to an
expected lifetime fraction~\shortcite{achatz2026}. The
\gls{pof} tradition that underpins lifetime modeling in
this domain combines Coffin--Manson estimates with Miner's rule to
handle variable stress
profiles~\shortcite{ceccarelli2019,barbagallo2021}, and
\gls{piml} frameworks have started to bridge it with data-driven
approaches~\shortcite{fassi2024}. From a broader \gls{phm}
standpoint, three persistent issues have been repeatedly flagged
in the literature: data-driven prognostics are typically evaluated
on training-distribution-matched test sets, the domain shift
between accelerated tests and field operation is seldom assessed,
and predictive uncertainty is rarely reported alongside the point
estimate~\shortcite{zio2022,fink2020}, and these gaps are confirmed
to remain open in our \gls{sic} domain~\shortcite{kumar2026}.

In this work we examine an accelerated dataset whose degradation
pattern departs from those most frequently analyzed in the
literature. The modules under test are 1.2~kV automotive \gls{sic}
half-bridges assembled with a sintered silver die-attach
technology, which suppresses solder degradation as a failure mode
and leaves wirebond fatigue as the dominant aging mechanism. On
the on-state voltage drop, this manifests as a continuous drift
driven by progressive crack propagation in the bond-pad
metallization, with rapid rises that occurs stocastically 
whenever a wire lifts off. Four stress conditions, obtained by
combining heating durations of 1.5~s and 30~s with different
load-current levels, all exhibit this profile, to get
module-to-module variability within each condition.

To handle this setting, we explore three complementary additions to
a standard data-driven baseline, one acting on the input, one on the
training loss, and one on the output. On the input side, the raw
indicator vector is augmented with physics-informed features that
summarize cumulative thermo-mechanical stress along the module
lifetime: the per-cycle junction-temperature swing and mean junction
temperature, their integrals over time, and a Miner-rule damage
term obtained from a per-cycle Coffin--Manson estimate. The
Miner-rule construction is the canonical physics-of-failure
aggregator for SiC under realistic mission
profiles~\shortcite{ceccarelli2019,barbagallo2021}; here it enters
the network as a learned input rather than as the prediction
itself. On the loss side, a monotonicity prior is added as a
gradient-penalty regularizer on the features that are physically
expected to grow with degradation in line with the loss-based
formulation used in our earlier work ~\shortcite{scarpa2025}. 
Compared with approaches that 
enforce monotonicity by construction, through custom activation 
functions inside the network~\shortcite{zhaoFink2025}, this
soft formulation introduces no inference overhead, remains
compatible with simple feed-forward networks, and matches the
constraints of embedded deployment. On the output side, the
network produces the parameters of a Student-$t$ distribution
rather than a single point estimate, so that location, scale, and
degrees-of-freedom are learned jointly and the heavy tails
accommodate the larger residuals that lift-off events produce in
the regression target.

The framework is evaluated under a four-fold cross-validation
stratified at the module level. Three neural architectures
(\gls{mlp}, \gls{cmlp}, \gls{node}) are trained with
different input-feature configurations: a baseline using only the
line current and the relative on-state voltage drop, and the same
set augmented with the cumulative features and the monotonicity
penalty. 
The combination of cumulative features and monotonicity
penalty reduces the \gls{mae} on the lifetime estimate by
up to $60\%$ relative to the baseline, at a
cross-fold standard deviation of $0.019$ on the health index that
holds also on the folds whose validation modules exhibit
lift-off. \Cref{sec:bg}
describes the failure physics and the indicator dataset.
\Cref{sec:framework} introduces the feature pipeline, the loss
formulation, and the output head. \Cref{sec:experimental_setup}
details the experimental protocol. \Cref{sec:results} reports the
comparative evaluation. \Cref{sec:discussion,sec:conclusion} close
with the discussion and the open questions.

% ============================================================
\section{Background and Problem Formulation}
\label{sec:bg}
% ============================================================

\subsection{Module and Power-Cycling Dataset}
\label{sec:dataset}

The dataset analyzed in this work consists of accelerated
power-cycling tests performed at Infineon Technologies on a
1.2~kV automotive \gls{sic} half-bridge module. Its sintered silver die-attach
removes solder fatigue at the chip--substrate interface as a
failure mode, leaving wirebond fatigue at the chip-side bond pads
as the dominant package-level mechanism. As thermal cycles
accumulate, mechanical stress drives crack propagation in the
bond-pad metallization, and individual wire lift-offs translate
into the rapid rises of the on-state voltage drop described in
\cref{sec:intro}.

Each \gls{dut} is repeatedly heated by a load current
$I_\text{load}$ flowing for a heating duration $t_\text{heat}$,
then cooled to the inlet coolant temperature over a duration
$t_\text{cool}$. During each cycle, the per-cycle on-state voltage
drop $V_{DS}$, the thermal resistance $R_\text{th}$, the
junction-temperature swing $\Delta T_\text{vj} = T_\text{vj,max} -
T_\text{vj,min}$, together with the load current and the
heating/cooling durations, are recorded. Following the \gls{aqg}~324
qualification standard~\shortcite{aqg324}, a module is considered
failed when at least one of
\begin{equation}
\frac{\Delta V_{DS}}{V_{DS,\text{nom}}} \geq 5\%, \qquad
\frac{\Delta R_\text{th}}{R_\text{th,nom}} \geq 20\%
\label{eq:eol_criterion}
\end{equation}
is reached, where $V_{DS,\text{nom}}$ and $R_\text{th,nom}$ are the
reference values measured on the pristine module. In the present
dataset, crossing of the $V_{DS}$ threshold is the primary
\gls{eol} event observed, while the thermal-resistance excursion
remains bounded throughout the test.

Twenty-four \glspl{dut} are distributed across four
operating-condition groups, summarized in
\cref{tab:operating_conditions}. The
groups span two heating-pulse durations (short, $t_\text{heat} =
1.5$~s, and long, $t_\text{heat} = 30$~s) and four load-current
levels, with the inlet coolant temperature trimmed
condition-by-condition to keep the junction-temperature peak close
to $T_\text{vj,max} \approx 175\,^\circ\text{C}$ across conditions.
The short-pulse groups compress the same heating energy into a
shorter cycle, inducing larger $\Delta T_\text{vj}$ excursions than
the long-pulse groups. The same power-cycling protocol on a
related solder-based \gls{sic} module has been previously
analyzed~\shortcite{olschewski2025}; the experimental framework is
therefore consistent across studies, and only the packaging
technology differs.

\begin{figure}[t]
\centering
\includegraphics[width=\linewidth]{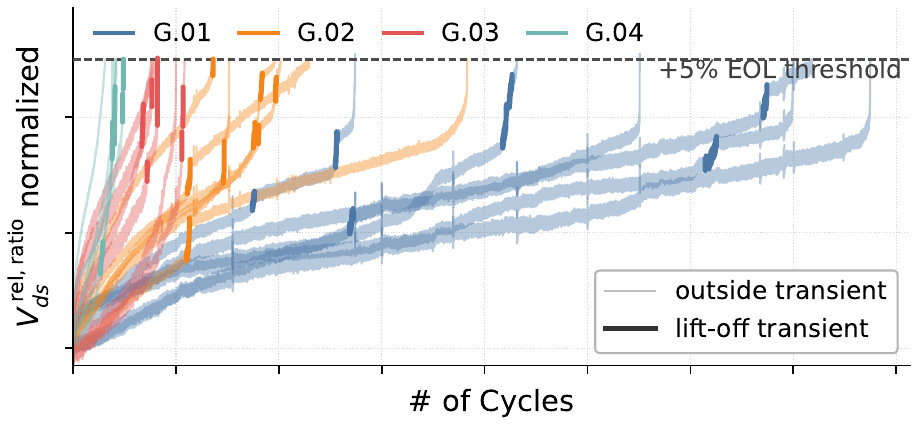}
\caption{Relative on-state voltage drop of each DUT along its
cycle counter. The light curves show the underlying drift; the
solid-colour overlays mark the cycle ranges annotated as wire
lift-off transients during data preprocessing. Numeric tick
labels are suppressed; the dashed line marks the $+5\%$ EOL
threshold.}
\label{fig:vds_traj}
\end{figure}

\Cref{fig:vds_traj} reports the relative on-state voltage
drop of each \gls{dut} along its cycle counter, with numeric tick
labels intentionally suppressed and only the $+5\%$ \gls{eol}
threshold marked. Most trajectories drift smoothly under the
continuous crack propagation in the bond-pad metallization; on top
of this drift, the cycle ranges in which wire lift-off transients
have been manually identified during data preprocessing are
overlaid in solid colour. Out of the $24$ \glspl{dut}, $19$
exhibit at least one such transient ($27$ events in total across
the four groups).

\begin{table}[t]
\centering
\caption{Operating conditions of the four power-cycling groups.
Current and temperatures are reported as mean $\pm$
standard deviation over all recorded cycles of the six DUTs of the
group, both switches pooled for $T_\text{vj,max}$; each group
contains $N = 6$ DUTs.}
\label{tab:operating_conditions}
\setlength{\tabcolsep}{3pt}
\begin{tabular}{lccccc}
\toprule
Group & $t_\text{heat}$ & $t_\text{cool}$ & $I_\text{load}$
      & $T_\text{vj,max}$ & $T_\text{inlet}$ \\
      & [s] & [s] & [A] & [$^\circ$C] & [$^\circ$C] \\
\midrule
G.01 & 1.5 & 3.5 & $694.5 \pm 0.05$ & $177.2 \pm 1.2$ & $73.3 \pm 0.09$ \\
G.02 & 1.5 & 3.5 & $752.6 \pm 0.05$ & $178.2 \pm 2.0$ & $53.6 \pm 0.02$ \\
G.03 & 30  & 30  & $669.5 \pm 0.07$ & $176.6 \pm 1.6$ & $74.1 \pm 0.17$ \\
G.04 & 30  & 30  & $776.7 \pm 0.07$ & $176.1 \pm 3.0$ & $33.1 \pm 0.06$ \\
\bottomrule
\end{tabular}
\end{table}
\subsection{Lifetime Modeling and Damage Aggregation}
\label{sec:lifetime}

\begin{figure}[t]
\centering
\includegraphics[width=\linewidth]{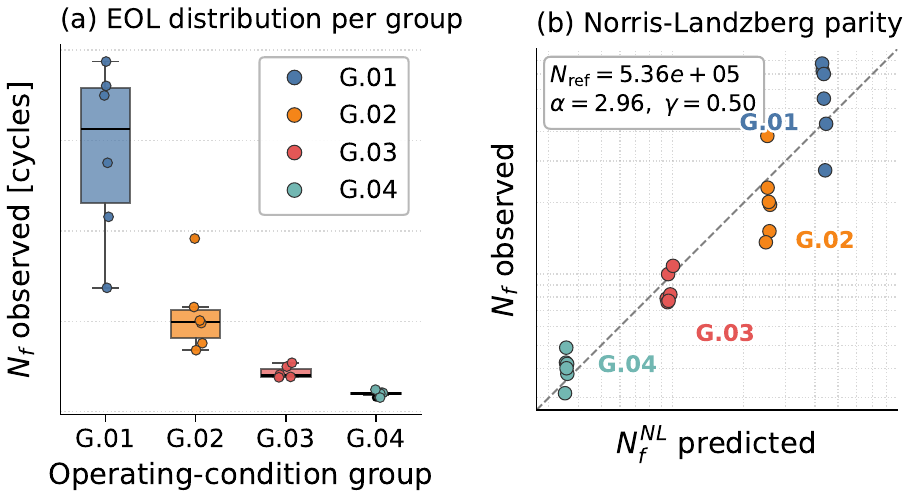}
\caption{(a) Empirical \gls{eol} distribution per
operating-condition group as per \cref{tab:miner_damage} on the dataset (six \glspl{dut} per
group). (b) Parity plot of observed and \gls{nl}-predicted cycles
to failure for the same \glspl{dut}; the dashed line marks ideal
agreement.}
\label{fig:eol_nl}
\end{figure}

The \gls{pof} framework that traditionally underpins lifetime
estimation for power modules combines a Coffin--Manson-type
estimate of cycles to failure under constant stress with Miner's
rule for aggregating variable stress
histories~\shortcite{ceccarelli2019,barbagallo2021}. For
wirebond-driven fatigue, we adopt the \gls{nl} parameterization,
which expresses the cycles to failure under a constant operating
point as
\begin{equation}
N_f^\text{NL}(\Delta T_\text{vj}, t_\text{heat}) =
N_\text{ref} \left(\frac{\Delta T_\text{vj}}{100}\right)^{-\alpha}
              t_\text{heat}^{-\gamma},
\label{eq:nl_model}
\end{equation}
where the three coefficients $N_\text{ref}$, $\alpha$, $\gamma$ are
fitted to the \gls{eol} data of the four groups in
\cref{tab:operating_conditions}. Log-linear least-squares
fitting on the 24 \glspl{dut} yields
\begin{equation}
N_\text{ref} = 5.32 \times 10^5, \quad
\alpha = 2.94, \quad
\gamma = 0.50,
\label{eq:nl_params}
\end{equation}
with $\Delta T_\text{vj}$ averaged over the early-stable cycle
window $[100, 600]$ to remove degradation-induced drift from the
operating-point estimate. \Cref{fig:eol_nl}(a) reports the
empirical \gls{eol} distribution per group as a boxplot, and
\cref{fig:eol_nl}(b) compares the observed cycles to failure
of each \gls{dut} with the \gls{nl} prediction evaluated
at~\cref{eq:nl_params}.

To aggregate variable stress histories at the \gls{dut} level,
Miner's rule is applied cycle-by-cycle. For a module that has
completed $n$ cycles under stress profile
$\{(\Delta T_{\text{vj},k}, t_{\text{heat},k})\}_{k=1}^{n}$, the
accumulated damage is
\begin{equation}
D(n) = \sum_{k=1}^{n}
\frac{1}{N_f^\text{NL}(\Delta T_{\text{vj},k}, t_{\text{heat},k})},
\label{eq:miner}
\end{equation}
with $D = 1$ corresponding to the predicted end of life. Evaluated
at the recorded \gls{eol} of each \gls{dut}, the Miner-damage
distribution is summarized in \cref{tab:miner_damage}.

\begin{table}[t]
\centering
\caption{Miner-rule damage~\cref{eq:miner} evaluated at the
recorded \gls{eol} of each \gls{dut}, reported per group and
aggregated.}
\label{tab:miner_damage}
\begin{tabular}{lcccc}
\toprule
Experiment & $n$ & $\overline{D}$ & $\sigma_D$ & CV \\
\midrule
G.01   & 6  & 1.323 & 0.419 & 32\% \\
G.02   & 6  & 0.855 & 0.324 & 38\% \\
G.03   & 6  & 0.898 & 0.116 & 13\% \\
G.04   & 6  & 1.132 & 0.146 & 13\% \\
\midrule
Global & 24 & 1.052 & 0.338 & 32\% \\
\bottomrule
\end{tabular}
\end{table}

The global aggregate $\overline{D} = 1.052$ confirms that the
\gls{nl} fit is on average well-centred on the dataset. The
per-group \gls{cv}, however, ranges from $13\%$ in the long-pulse
groups (G.03, G.04) to $38\%$ in the
short-pulse group G.02,
signalling that a substantial fraction of the module-to-module
variability remains unexplained at the operating-condition level.

\subsection{Problem Formulation}
\label{sec:problem_formulation}

The \gls{nl}/Miner framework provides a bulk estimate of the
cycles to failure at the population level and a scalar damage
index $D$ at the \gls{dut} level, both convenient for
accelerated-test characterisation. It does not provide a per-cycle
estimate of the current health state, nor a quantification of
predictive uncertainty around that estimate, nor a mechanism to
track module-specific deviations from the empirical fit. These
three quantities are the natural targets of an in-service \gls{cm}
tool.

We therefore introduce a dimensionless health index
\begin{equation}
r(n) = \frac{n}{N_f^\text{EOL}}, \qquad r \in [0,\, 1+\varepsilon],
\label{eq:health_index}
\end{equation}
where $n$ is the current cycle count and $N_f^\text{EOL}$ is the
cycle index at which the \gls{eol}
criterion~\cref{eq:eol_criterion} is first satisfied. By
construction, $r = 0$ corresponds to a pristine module, $r = 1$ to
the \gls{eol} event, and a small overshoot $\varepsilon > 0$ is
allowed to absorb the discretisation of \gls{eol} detection. The
\gls{cm} task addressed in the next section is to estimate $r(n)$
from the per-cycle measurements available at cycle $n$, together
with a predictive distribution that captures both module-to-module
residual variability and the larger residuals produced by
individual lift-off events.

% ============================================================
\section{Proposed Framework}
\label{sec:framework}
% ============================================================

\subsection{Network Architectures}
\label{sec:architectures}

We compare three different neural architectures that differ in
how the cycle history is exposed to the input: a snapshot
\gls{mlp}, which uses only the current measurement; a contextual
\gls{mlp} (\gls{cmlp}), which appends a short window of past
on-state voltage values; and a \gls{node}, which carries a hidden
state across cycles. All three map a per-cycle input vector $x_n
\in \mathbb{R}^d$ to the parameters of the predictive distribution
introduced in \cref{sec:output}.

The \gls{mlp} is the simplest of the three. It treats every cycle
independently, mapping a single snapshot $x_n$ through $L$
fully-connected layers with non-linear activation~\shortcite{rumelhart1986},
\begin{equation}
h_1 = \phi(W_1 x_n + b_1), \quad
h_\ell = \phi(W_\ell\, h_{\ell-1} + b_\ell),
\label{eq:mlp}
\end{equation}
with the output read from the last layer provided as
$[\mu_n, \tilde\sigma_n, \tilde\nu_n] = W_\text{out}\, h_L +
b_\text{out}$. The stateless formulation is straightforward to
train and fits comfortably within the parameter budget of an
embedded target, but it is blind to the cycle history: any signal
that requires comparing the current snapshot with the past must
already be encoded in $x_n$.

The contextual \gls{mlp} extends the input vector with a size limited
rolling window of past on-state voltage measurements,
\begin{equation}
\tilde x_n = \bigl[\,x_n,\ V_{DS,n-K+1}, \dots, V_{DS,n}\bigr]
\in \mathbb{R}^{d+K},
\label{eq:cmlp}
\end{equation}
and feeds $\tilde x_n$ through an \gls{mlp} with the same forward
form as~\cref{eq:mlp}. This restores access to a finite local
context without introducing a recurrent state. The trade-off is in
the choice of $K$, which is fixed by hand and cannot be learned:
anything outside the window is invisible to the model.

The \gls{node} replaces the discrete stacking of layers with a
continuous-time vector field on a hidden
state~\shortcite{chen2018node},
\begin{equation}
\frac{d h(t)}{d t} = f_\theta\bigl(h(t),\, x(t)\bigr),
\label{eq:node_ode}
\end{equation}
which is integrated forward in time by an explicit Euler step,
\begin{equation}
h_{n+1} = h_n + \Delta t_n \cdot f_\theta(h_n,\, x_n),
\label{eq:node_discrete}
\end{equation}
and produces the output by reading the hidden state,
$[\mu_n, \tilde\sigma_n, \tilde\nu_n] = W_\text{out}\, h_n +
b_\text{out}$. The step size $\Delta t_n$ is set proportional to
the per-cycle heating duration $t_{\text{heat},n}$, so that the
dynamics learned by $f_\theta$ are decoupled from the
operating-point-dependent cycle duration; the details of this
scaling are deferred to \cref{sec:experimental_setup}.
Compared with the previous two architectures, the \gls{node}
carries unbounded temporal memory in its hidden state and is the
only one whose recurrence is conditioned explicitly on
$t_\text{heat}$, at the cost of harder hyper-parameter tuning.
Training is performed on subsequences rather than on the full
trajectory, so the hidden state at each subsequence start is set
to zero; at inference, however, the state at any cycle results
from the integration of the dynamics from cycle zero onward, and
the zero-initialised training state is therefore inconsistent
with the state actually reached during inference. To reduce this
mismatch, we periodically run a full-trajectory simulation on the
training set and use the resulting hidden state at each
subsequence boundary as the new initialisation, in a manner
analogous to~\shortcite{scarpa2025}. The refresh frequency and
the subsequence spacing are reported in
\cref{sec:experimental_setup}.

\subsection{Probabilistic Output Head}
\label{sec:output}
In place of a Gaussian point-estimate trained against the mean
squared error, the network produces the parameters of a
Student-$t$ distribution over the health index $r$,
\begin{equation}
p(r \mid x_n) = \mathcal{T}\bigl(\mu_n,\ \sigma_n,\ \nu_n\bigr),
\label{eq:student_t}
\end{equation}
where the location $\mu_n$, the scale $\sigma_n > 0$ and the
degrees-of-freedom $\nu_n > 0$ are all produced by the network
for each cycle. Written out, the density is
\begin{equation}
p(r \mid x_n) =
\frac{\Gamma\bigl(\frac{\nu_n + 1}{2}\bigr)}
     {\Gamma\bigl(\frac{\nu_n}{2}\bigr)\sqrt{\pi \nu_n}\,\sigma_n}
\left[
1 + \frac{1}{\nu_n}
\left(\frac{r - \mu_n}{\sigma_n}\right)^{\!2}
\right]^{-\frac{\nu_n + 1}{2}} ,
\label{eq:student_t_pdf}
\end{equation}
which recovers the Gaussian in the limit
$\nu_n \to \infty$ and grows heavier tails as $\nu_n$ decreases.
The three parameters are learned end-to-end
through the negative log-likelihood of~\cref{eq:student_t},
\begin{equation}
\mathcal{L}_\text{NLL} = -\frac{1}{N} \sum_{n=1}^{N} \log p(r_n
\mid x_n).
\label{eq:nll}
\end{equation}
A Gaussian likelihood, equivalent to the mean squared error up
to constants, penalises large residuals quadratically. In our
setting the cycles in which an individual wire lift-off occurs
are exactly the cycles where the residual on $r$ is largest, so a
quadratic loss amplifies the corresponding gradient contributions
and tends to destabilise training. The Student-$t$ replaces the
quadratic tail with a $\log\bigl(1 + (r - \mu)^2/(\nu\sigma^2)\bigr)$
tail whose curvature decreases as $\nu$ shrinks. The learned
$\nu_n$ then acts as a per-cycle reliability measure rather than a
fixed hyper-parameter, and the high-residual cycles contribute
proportionally less to the gradient than they would under a
Gaussian assumption.

\subsection{Monotonicity Prior as a Soft Constraint}
\label{sec:monotonicity}

The health index $r(n) = n / N_f^\text{EOL}$ is by construction
non-decreasing in $n$, and the cumulative features introduced in
\cref{sec:problem_formulation} are non-decreasing in the
relevant inputs as well. We encode this physical prior as a
gradient-penalty regulariser added to the training objective. For
the stateless architectures (\gls{mlp} and \gls{cmlp}), the
constraint is imposed pointwise as a non-negativity penalty on
the partial derivative of the central estimate with respect to
each cumulative input,
\begin{equation}
\mathcal{L}_\text{mono}^\text{pw} = \frac{1}{N} \sum_{n=1}^{N}
\sum_{i \in \mathcal{I}_\text{cum}}
\Bigl[\, -\frac{\partial \mu_n}{\partial x_{n,i}} \Bigr]_+^{2},
\label{eq:mono_pw}
\end{equation}
where $[\cdot]_+$ denotes the positive part and
$\mathcal{I}_\text{cum}$ collects the indices of the physically
monotone features. For the recurrent architecture (\gls{node}),
the same prior can also be imposed temporally on the predicted trajectory,
\begin{equation}
\mathcal{L}_\text{mono}^\text{t} = \frac{1}{N-1} \sum_{n=1}^{N-1}
\Bigl[\, -(\mu_{n+1} - \mu_n) \Bigr]_+^{2}.
\label{eq:mono_t}
\end{equation}
In both cases the training objective combines the two terms with
a non-negative weight $\lambda$,
\begin{equation}
\mathcal{L} = \mathcal{L}_\text{NLL} + \lambda \,
\mathcal{L}_\text{mono},
\label{eq:total_loss}
\end{equation}
and the prior acts on the loss only at training time, so the
inference graph is left unchanged and no overhead is added at
deployment.

% ============================================================
\section{Experimental Setup}
\label{sec:experimental_setup}
% ============================================================

\subsection{Input Features}
\label{sec:features}

Each architecture is fed a per-cycle vector $x_n$ built from two
groups of features. The \emph{base} group contains the line
current and the relative on-state voltage drop,
\begin{equation}
x_n^\text{base} = \bigl[I_{\text{load},n},\ V_{DS,n}^{\,\text{rel}}\bigr]
\in \mathbb{R}^{2},
\label{eq:base_features}
\end{equation}
which are the two quantities measured at every cycle in the
power-cycling test. The \emph{cumulative} group collects four
physics-informed quantities that summarise the stress history up
to cycle $n$,
\begin{equation}
\begin{aligned}
S_{T_j}(n) &= \textstyle\sum_{k=1}^{n} \int_{\text{t}_k}
              T_{\text{vj,ss}}(\tau)\, d\tau,
\\[3pt]
S_{\Delta T_j}(n) &= \textstyle\sum_{k=1}^{n} \Delta T_{\text{vj},k},
\\[3pt]
S_{I}(n) &= \textstyle\sum_{k=1}^{n} \int_{\text{t}_k}
            I_{\text{load}}(\tau)\, d\tau,
\\[3pt]
D(n) &= \text{Miner damage of}~\cref{eq:miner},
\end{aligned}
\label{eq:cum_features}
\end{equation}
all of which are by construction non-decreasing in $n$ and form
the set $\mathcal{I}_\text{cum}$ on which the pointwise
monotonicity penalty of \cref{eq:mono_pw} acts. The baseline
configuration uses only $x_n^\text{base}$; the full configuration
appends~\cref{eq:cum_features}, so that
$x_n \in \mathbb{R}^{6}$.

\subsection{Network Sizes and Training}
\label{sec:training}

The \gls{mlp} and \gls{cmlp} share a three-layer feed-forward
network with ReLU activation and a hidden width of about $15$
units per layer; the last layer outputs the three Student-$t$
parameters $(\mu_n,\tilde\sigma_n,\tilde\nu_n)$. The \gls{cmlp}
appends $K = 10$ past values of $V_{DS}^{\,\text{rel}}$ to the
per-cycle input as in~\cref{eq:cmlp}. The \gls{node} uses a
hidden state of comparable dimension and a two-layer \gls{mlp} as
the vector field $f_\theta$. The hidden widths of all three
networks are chosen so that, on the full feature set, the total
parameter count remains below $10^3$ trainable weights, keeping
the three models within the same order of magnitude so that the
comparison isolates the effect of the temporal context exposed to
the input from the effect of model size. The
integration step in \cref{eq:node_discrete} is
\begin{equation}
\Delta t_n = \frac{t_{\text{heat},n}}{\textsc{dt}_\text{scale}},
\label{eq:dt_scale}
\end{equation}
which is the explicit mechanism that decouples the dynamics
learned by $f_\theta$ from the per-group cycle duration: under the
power-cycling protocol of \cref{tab:operating_conditions} the
heating duration varies by a factor of $20$ between the short and
the long groups, so that a constant integration step would force
the network to absorb this variability into the weights rather
than into the recurrence. We set $\textsc{dt}_\text{scale} =
1000$, the same value across all groups, so that $\Delta t_n$
remains in the $[10^{-3},\, 10^{-2}]$ range throughout the
dataset.

The Student-$t$ scale and degrees-of-freedom are passed through a
softplus to keep them strictly positive,
$\sigma_n = \text{softplus}(\tilde\sigma_n) + \sigma_\text{floor}$
and $\nu_n = \text{softplus}(\tilde\nu_n) + \nu_\text{min}$, with
$\sigma_\text{floor} = 0.02$ to prevent vanishing scale and
$\nu_\text{min} = 2$ to keep the second moment of the heavy-tail
likelihood finite. Optimisation uses Adam with a learning rate of
$10^{-3}$ scheduled piecewise with a $0.1$ decay, gradient norm
clipped at $1.0$, and a monotonicity weight of $\lambda = 10$ when
the penalty is active. All three architectures are trained
for $500$ epochs with a batch size of $2{,}048$ samples; for the
\gls{node} the batch is composed of subsequences of $5{,}000$
cycles drawn from the training \glspl{dut} with a stride of $500$
cycles, while the \gls{mlp} and \gls{cmlp} are fed batches of
independent per-cycle snapshots. The full-trajectory simulation
that refreshes the hidden-state cache at each subsequence boundary
is re-run every $30$ epochs. All networks are fed a $10\times$
subsampled version of the cycle counter to keep the per-epoch
compute budget manageable on embedded-class hardware.

\subsection{Cross-Validation Protocol and Baseline}
\label{sec:cv}

The $24$ \glspl{dut} are split with a four-fold cross-validation
stratified at the operating-condition group level: each fold
reserves one \gls{dut} per group for validation and uses the
remaining $20$ \glspl{dut} for training. The folds are mutually
exclusive at the \gls{dut} level, so every \gls{dut} is used
exactly once as validation across the four folds. Performance is
reported as mean and standard deviation across folds.

The reference against which the proposed configurations are
compared is the parameter-free estimator
\begin{equation}
\hat r_n^{\,\text{base}} = 20\, V_{DS,n}^{\,\text{rel}},
\label{eq:trivial_baseline}
\end{equation}
obtained by inverting the $5\%$ \gls{aqg}~324 criterion
of~\cref{eq:eol_criterion} into a linear map from
$V_{DS}^{\,\text{rel}} \in [0, 0.05]$ to $r \in [0, 1]$. This
estimator requires no training and uses only the indicator that
the qualification standard already monitors. Any improvement of
the proposed networks over~\cref{eq:trivial_baseline} therefore
quantifies the information contributed by the cumulative features
and the monotonicity prior on top of the raw $V_{DS}$ trace.

\subsection{Evaluation Metrics}
\label{sec:metrics}

Four metrics are reported per fold, each computed over all
validation cycles. The first two assess the central estimate
$\hat r_n = \mu_n$ against the true health index $r_n$,
\begin{align}
\text{MAE} & = \frac{1}{N} \sum_{n=1}^{N} |\hat r_n - r_n|,
 \label{eq:mae} \\
R^2 & = 1 - \frac{\sum_n (r_n - \hat r_n)^2}
              {\sum_n (r_n - \bar r)^2}
\label{eq:r2}
\end{align}
where $\bar r$ is the mean of $r$ on the validation set. The
remaining two are standard prognostic metrics that weigh the
estimate against the residual life~\shortcite{saxena2010},
\begin{equation}
\alpha\text{-acc} = \frac{1}{N} \sum_{n=1}^{N}
\mathbb{1}\bigl[\,|\hat r_n - r_n| \leq \alpha\,(1 - r_n)\,\bigr],
\label{eq:alpha_acc}
\end{equation}
which measures the fraction of cycles inside an $\alpha$-cone
that narrows toward the \gls{eol} (we use $\alpha = 0.2$), and the
relative accuracy
\begin{equation}
\text{RA} = 1 - \frac{|\hat r_n - r_n|}{r_n},
\label{eq:ra}
\end{equation}
averaged over the cycles in which $r_n \geq 0.1$ to keep the
denominator away from zero in the pristine region.

% ============================================================
\section{Results}
\label{sec:results}
% ============================================================

\subsection{Aggregate Cross-Validation Performance}
\label{sec:results_aggregate}

The four metrics of \cref{eq:mae,eq:r2,eq:alpha_acc,eq:ra},
aggregated as mean and standard deviation across the four
cross-validation folds, are reported in \cref{tab:results}. The
trivial baseline of \cref{eq:trivial_baseline} sets a non-trivial
floor that any learned estimator must clear. Training on the base
features alone --- the telemetric signals available from the
inverter without any explicit history encoding --- yields only a
modest \gls{mae} improvement and keeps $\alpha$-accuracy
essentially at the level of the trivial reference. Appending the
cumulative features of \cref{eq:cum_features}, which encode the
stress history experienced by the module up to cycle $n$,
translates the input into a sharper estimate: the gap closes on
every metric and the recurrent \gls{node} becomes consistently
best. The monotonicity penalty of \cref{eq:mono_pw} adds a smaller
increment --- a few percentage points on $\alpha$-accuracy and a
fraction of a percent on \gls{mae} --- consistent with the
hypothesis that the cumulative features themselves already encode
a substantial degree of physical monotonicity, partly carrying out
the role of the explicit regularizer.

Beyond the best mean performance, the \gls{node} also exhibits
the lowest fold-to-fold variance on every metric, as shown by the
standard deviation columns of \cref{tab:results}. This stability
is consistent with the recurrent hidden state acting as an
implicit memory of the trajectory, which absorbs the
cumulative-damage signal across regular cycles as well as across
anomalous events such as the wirebond lift-off transients that
disrupt $V_{DS,n}^{\,\text{rel}}$ on individual \glspl{dut}.

\begin{table*}[!t]
\centering
\caption{Cross-validation metrics on the dataset: mean
$\pm$ standard deviation across the four \gls{dut}-stratified
folds described in \cref{sec:cv}. Arrows indicate the direction
of improvement; the smaller line below each value reports the
relative improvement over the trivial baseline (positive $=$
better, irrespective of the metric's direction). The
best-performing configuration is highlighted in bold.}
\label{tab:results}
\begin{tabular}{llcccc}
\toprule
Architecture & Features & MAE $\downarrow$ & $R^2$ $\uparrow$ & $\alpha$-acc $\uparrow$ & RA $\uparrow$ \\
\midrule
\multirow{2}{*}{$20\,V_{DS}^{\,\text{rel}}$ (baseline)} & \multirow{2}{*}{---} & $0.120\!\pm\!0.013$ & $0.738\!\pm\!0.078$ & $0.444\!\pm\!0.037$ & $0.716\!\pm\!0.017$ \\
 & & {\scriptsize ---} & {\scriptsize ---} & {\scriptsize ---} & {\scriptsize ---} \\
\midrule
\multirow{6}{*}{\gls{mlp}} & \multirow{2}{*}{base} & $0.093\!\pm\!0.020$ & $0.816\!\pm\!0.049$ & $0.489\!\pm\!0.039$ & $0.695\!\pm\!0.015$ \\
 & & {\scriptsize (+23\%)} & {\scriptsize (+11\%)} & {\scriptsize (+10\%)} & {\scriptsize (-3\%)} \\
\cmidrule(l){2-6}
 & \multirow{2}{*}{cum (no penalty)} & $0.073\!\pm\!0.047$ & $0.846\!\pm\!0.171$ & $0.667\!\pm\!0.211$ & $0.836\!\pm\!0.083$ \\
 & & {\scriptsize (+40\%)} & {\scriptsize (+15\%)} & {\scriptsize (+50\%)} & {\scriptsize (+17\%)} \\
\cmidrule(l){2-6}
 & \multirow{2}{*}{cum + monotonicity} & $0.072\!\pm\!0.048$ & $0.850\!\pm\!0.173$ & $0.673\!\pm\!0.219$ & $0.839\!\pm\!0.085$ \\
 & & {\scriptsize (+40\%)} & {\scriptsize (+15\%)} & {\scriptsize (+51\%)} & {\scriptsize (+17\%)} \\
\midrule
\multirow{6}{*}{\gls{cmlp}} & \multirow{2}{*}{base} & $0.097\!\pm\!0.030$ & $0.723\!\pm\!0.193$ & $0.413\!\pm\!0.174$ & $0.653\!\pm\!0.071$ \\
 & & {\scriptsize (+19\%)} & {\scriptsize (-2\%)} & {\scriptsize (-7\%)} & {\scriptsize (-9\%)} \\
\cmidrule(l){2-6}
 & \multirow{2}{*}{cum (no penalty)} & $0.068\!\pm\!0.035$ & $0.893\!\pm\!0.072$ & $0.629\!\pm\!0.183$ & $0.817\!\pm\!0.065$ \\
 & & {\scriptsize (+43\%)} & {\scriptsize (+21\%)} & {\scriptsize (+42\%)} & {\scriptsize (+14\%)} \\
\cmidrule(l){2-6}
 & \multirow{2}{*}{cum + monotonicity} & $0.063\!\pm\!0.028$ & $0.911\!\pm\!0.049$ & $0.662\!\pm\!0.166$ & $0.819\!\pm\!0.061$ \\
 & & {\scriptsize (+48\%)} & {\scriptsize (+23\%)} & {\scriptsize (+49\%)} & {\scriptsize (+14\%)} \\
\midrule
\multirow{6}{*}{\gls{node}} & \multirow{2}{*}{base} & $0.077\!\pm\!0.010$ & $0.850\!\pm\!0.026$ & $0.493\!\pm\!0.081$ & $0.717\!\pm\!0.018$ \\
 & & {\scriptsize (+36\%)} & {\scriptsize (+15\%)} & {\scriptsize (+11\%)} & {\scriptsize (+0\%)} \\
\cmidrule(l){2-6}
 & \multirow{2}{*}{cum (no penalty)} & $0.050\!\pm\!0.019$ & $0.944\!\pm\!0.028$ & $0.768\!\pm\!0.111$ & $0.860\!\pm\!0.043$ \\
 & & {\scriptsize (+58\%)} & {\scriptsize (+28\%)} & {\scriptsize (+73\%)} & {\scriptsize (+20\%)} \\
\cmidrule(l){2-6}
 & \multirow{2}{*}{\textbf{cum + monotonicity}} & \textbf{$0.049\!\pm\!0.019$} & \textbf{$0.947\!\pm\!0.029$} & \textbf{$0.792\!\pm\!0.126$} & \textbf{$0.859\!\pm\!0.045$} \\
 & & {\scriptsize \textbf{(+59\%)}} & {\scriptsize \textbf{(+28\%)}} & {\scriptsize \textbf{(+78\%)}} & {\scriptsize \textbf{(+20\%)}} \\
\bottomrule
\end{tabular}

\end{table*}

\subsection{Health Index and Relative Accuracy Across Life}
\label{sec:results_alphara}

\begin{figure*}[!t]
\centering
\includegraphics[width=\textwidth]{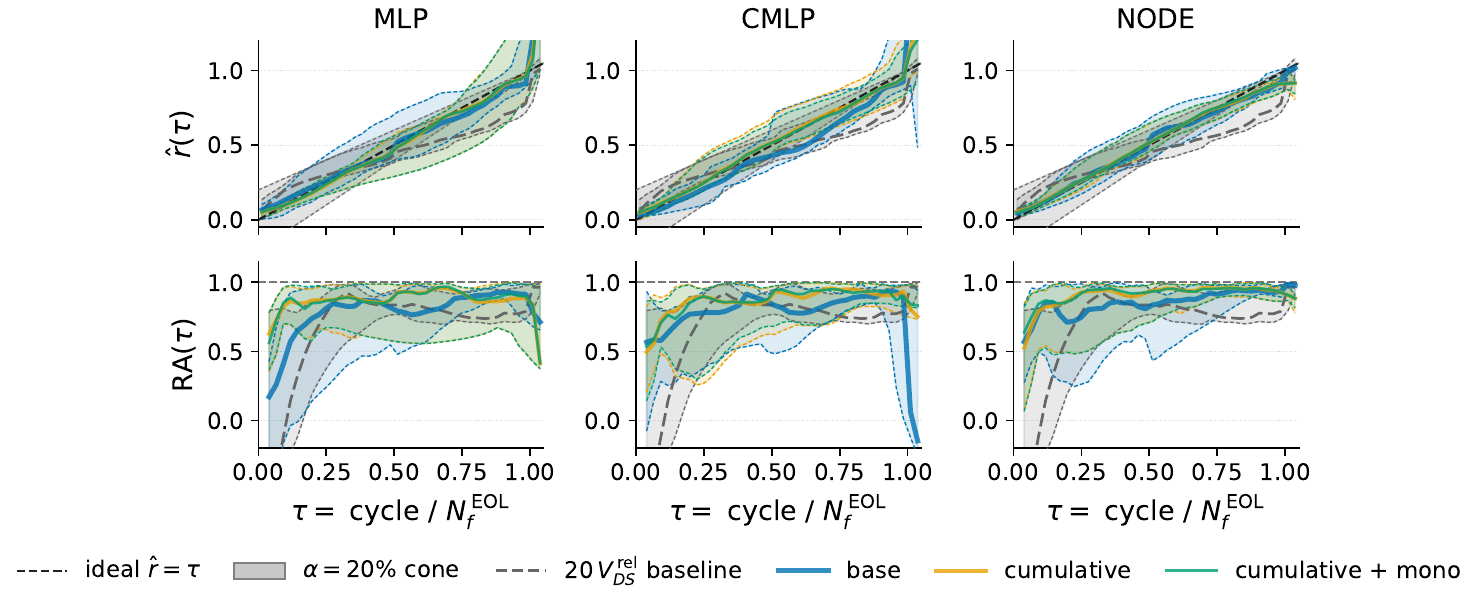}
\caption{Aggregate validation behaviour along the normalised life
$\tau = n / N_f^\text{EOL}$. \emph{Top:} predicted health index
$\hat r(\tau)$ with the ideal $\hat r = \tau$ diagonal and the
$\alpha = 20\%$ cone of \cref{eq:alpha_acc}. \emph{Bottom:}
relative accuracy $\text{RA}(\tau)$ of \cref{eq:ra}. Curves are
the cross-\gls{dut} median at each $\tau$ bin; dashed envelopes
mark the $10$th and $90$th percentiles over the validation
\glspl{dut} pooled across the four folds of \cref{sec:cv}. The
parameter-free baseline of \cref{eq:trivial_baseline} is drawn in
gray as reference in every panel.}
\label{fig:alpha_ra_grid}
\end{figure*}

The aggregate behaviour of the four series along the life fraction
$\tau = n / N_f^\text{EOL}$ is reported in
\cref{fig:alpha_ra_grid}, with one column per architecture and one
row per metric. The top row overlays the ideal $\hat r = \tau$
diagonal and the $\alpha = 20\%$ cone of \cref{eq:alpha_acc}; the
bottom row marks the $\text{RA} = 1$ ideal.

The trivial baseline, drawn in gray in all six panels, captures
the slow $V_{DS}$ drift correctly in mid-life but lags below the
diagonal for the first three quarters of the trajectory and only
catches up close to the \gls{eol} threshold. As a direct
consequence its $\text{RA}$ dives below zero for $\tau \in
[0.05,\, 0.15]$: in this regime $V_{DS,n}^{\,\text{rel}} \approx
0$ on most modules, so $\hat r$ is comparable in magnitude to the
absolute error itself and the denominator of \cref{eq:ra} is too
small to absorb it. This regime is precisely where a learned
estimator has the most to add: from $\tau = 0.2$ onward all three
architectures stay above $\text{RA} = 0.8$ on the median curve,
while the baseline only recovers to $\text{RA} \approx 0.7$ around
$\tau = 0.5$.

The base configuration (blue curve) is the one in which the choice
of architecture matters most. On the \gls{mlp} panel the base
median strays visibly outside the cone for $\tau \lesssim 0.2$ and
shows the widest $10$-$90$ band of the three columns, consistent
with the largest cross-fold variance reported in
\cref{tab:results}. The \gls{cmlp} corrects part of this through
the past-$V_{DS}$ context window; the \gls{node} closes most of
the gap through the recurrent state alone, with the base median
already inside the cone almost everywhere.

The cumulative variants (orange and green) compress these
differences across architectures: once $S_{T_j}$, $S_{\Delta T_j}$,
$S_I$ and $D$ are available, the three median curves essentially
overlap. The monotonicity penalty (green vs.\ orange) brings a
slight visible tightening of the band on the \gls{cmlp} and
\gls{node} panels for $\tau > 0.3$ but is barely distinguishable
on the \gls{mlp}. The bands of the \gls{node} cumulative variants
are the narrowest among the nine network configurations,
consistent with the lowest cross-fold variance of
\cref{tab:results}: the recurrent architecture pairs the best
central tendency with the most consistent inter-\gls{dut}
response.

\subsection{Predictive Uncertainty Calibration}
\label{sec:results_uncertainty}

We check whether the Student-$t$ output of \cref{eq:student_t}
delivers calibrated uncertainty. For each validation cycle we form the
Student-$t$ central interval at nominal coverage $p \in
\{68.27\%,\,95.45\%\}$ from the predicted $(\sigma_n,\nu_n)$ and count
the fraction of cycles in which the true $r_n$ falls inside that
interval. \Cref{tab:uncertainty} reports the result. The base
\gls{node} is undercovered at both levels: with only the raw
electrical inputs the network reports tighter bands than the
residuals justify. Adding the cumulative features
of~\cref{eq:cum_features} brings coverage within two percentage
points of nominal and shrinks the median predictive scale
$\tilde{\sigma}$ by roughly $37\%$. The monotonicity penalty leaves
the calibration essentially unchanged, in line with the small effect
it already had on the point-estimate metrics of \cref{tab:results}.
The median $\nu_n$ sits around six in all three variants and $\nu_n
< 10$ on every validation cycle: the heavy tails are always active
and never collapse to a Gaussian limit.

\begin{figure}[t]
\centering
\includegraphics[width=\linewidth]{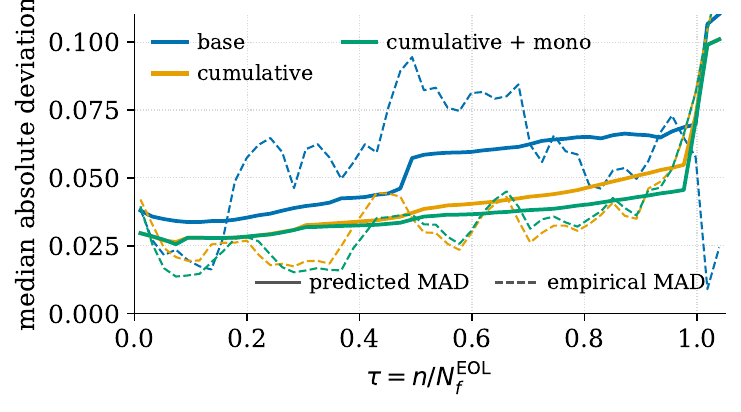}
\caption{Predicted versus empirical median absolute deviation along
the life fraction $\tau = n / N_f^\text{EOL}$ for the three \gls{node}
variants. The predicted curves (solid) are
$\sigma_n \cdot t^{-1}_{0.75,\,\nu_n}$, i.e.\ each $\sigma_n$ rescaled
by the standardised Student-$t$ median quantile so that it can be
compared directly to a deviation measure; the empirical curves
(dashed) are the cross-\gls{dut} median of $|r-\mu|$ binned at the
same resolution and pooled across the four folds. The cumulative
variants track their empirical counterparts within one bin width over
the central $\tau$ range; the base variant sits below its empirical
counterpart, consistent with the under-coverage reported
in~\cref{tab:uncertainty}.}
\label{fig:uncertainty}
\end{figure}

\Cref{fig:uncertainty} resolves the same comparison locally along the
life fraction $\tau$. The deviation-size statistic adopted at each
$\tau$ is the median absolute deviation (MAD), measured from the data
as the cross-\gls{dut} median of the residual,
\begin{equation}
\mathrm{MAD}^{\text{emp}}_n = \mathrm{med}\bigl|\,r_n - \mu_n\,\bigr|,
\label{eq:mad_emp}
\end{equation}
and predicted by the model as the analytic MAD of the Student-$t$
distribution at cycle $n$,
\begin{equation}
\mathrm{MAD}^{\text{pred}}_n
  = \sigma_n \cdot t^{-1}_{0.75,\,\nu_n},
\label{eq:mad_pred}
\end{equation}
where $t^{-1}_{0.75,\nu}$ is the standardised Student-$t$ median
quantile that rescales the scale parameter $\sigma_n$ into a deviation
and brings the two quantities onto the same axis. For the base
configuration the predicted MAD sits visibly below its empirical
counterpart over most of the trajectory: the network is not only
larger in absolute residual than the cumulative variants but also
overconfident on that residual, declaring a tighter spread than the
data actually exhibit. The cumulative variant brings the residual $\mathrm{MAE}$ from
$7.7\%$ to $5.0\%$ of the unit lifetime
range (\cref{tab:results}), a $35\%$ relative reduction, and at the
same time aligns its predicted MAD with the empirical one, so the gain on
accuracy and the gain on calibrated uncertainty come together. 
The cumulative with monotonicity
variant tightens this picture a step further, with the predicted MAD
slightly below the cumulative variant for most of $\tau$ without
losing the calibration alignment, consistent with the marginal
accuracy gain already observed in~\cref{tab:results}. The Student-$t$
output is then a usable per-cycle uncertainty signal in its own right,
and might be an input for the probabilistic prognostic metrics flagged
as missing in~\cref{sec:discussion}.

\begin{table}[t]
\centering
\caption{Predictive uncertainty calibration of the \gls{node} on the
validation cycles of the four-fold cross-validation. ``Cov.\ $p\%$''
is the empirical fraction of cycles inside the Student-$t$ central
interval at nominal coverage $p\%$, computed per-cycle using the
predicted $(\sigma_n,\nu_n)$; the matching nominal values are
$68.27\%$ and $95.45\%$. Median $\nu$ and the fraction of cycles with
$\nu_n < 10$ summarise the use of the heavy-tail regime;
$\tilde\sigma$ is the median predictive scale in percent of the unit
lifetime range.}
\label{tab:uncertainty}
\footnotesize
\setlength{\tabcolsep}{4pt}
\begin{tabular}{l c c c c c}
\toprule
NODE variant & $68\%$ cov.\ & $95\%$ cov.\ & $\widetilde{\nu}$ & $\nu{<}10$ & $\widetilde{\sigma}$ \\
 & [\%] & [\%] & & [\%] & [\%] \\
\midrule
base & 66.1 & 88.3 & 5.0 & 100.0 & 7.28 \\
cumulative & 70.7 & 96.6 & 6.5 & 100.0 & 4.61 \\
cumulative + mono & 70.9 & 97.4 & 6.6 & 100.0 & 4.58 \\
\bottomrule
\end{tabular}
\end{table}

% ============================================================
\section{Discussion}
\label{sec:discussion}
% ============================================================

The cross-validation results support an observation that
cuts across the three architectures: appending the cumulative
features of \cref{eq:cum_features} provides a net positive
contribution on every metric of \cref{tab:results}, and the
choice of architecture then modulates how much each network makes
of this enrichment. The monotonicity penalty adds a smaller,
architecture-dependent bump, primarily on \gls{mae} and $\alpha$-accuracy. The
broader reading is that, when degradation is dominated by a slow
accumulation of stress, encoding that history at the feature
level is a robust way to anchor the predictor to the physics; how
much benefit each architecture extracts from this anchor is then
a function of how well it combines the instantaneous history
estimate with the temporal context it provides on its own.

A few aspects of the experimental setting frame the scope of
these results. The protocol of \cref{tab:operating_conditions}
exercises the heating-pulse duration $t_\text{heat}$ at the two
levels prescribed by the \gls{aqg}~$324$ recipe, on disjoint
groups of \glspl{dut} and without mixing them within a single
trajectory. The integration step of \cref{eq:dt_scale} is
designed precisely to absorb such a mixing, and the consistent
behaviour of the \gls{node} across the two operating groups
suggests that the mechanism does carry useful information.
Verifying this directly would require dedicated campaigns with
$t_\text{heat}$ varying within a single \gls{dut}, in profiles
closer to the realistic mission of a traction inverter.

The metrics of \cref{sec:metrics} are point-estimate metrics; they
do not use the predictive distribution of the Student-$t$ head. We
rely on them because they are consolidated in the prognostics
literature~\shortcite{fink2020} and informative on different stages
of the \gls{sic} health-state estimate: \gls{mae} and $R^2$ summarise
the global residual, while $\alpha$-accuracy and RA emphasise the
late-life region where the tolerance shrinks and the denominator of
the relative error is small. A separate calibration analysis is
reported in~\cref{sec:results_uncertainty}. The general recommendation
to report uncertainty alongside a point estimate is also raised in
the same review; the $(\sigma_n, \nu_n)$
of~\cref{tab:uncertainty,fig:uncertainty} is the form adopted here.

The predictive scale $\sigma_n$ aggregates aleatoric and epistemic
contributions; the present training setup does not separate them.
With twenty-four \glspl{dut} the data budget for an explicit
epistemic spread is limited, and the $\sigma_n$ reported
in~\cref{sec:results_uncertainty} is dominated by the aleatoric
component conditional on the input. A future development could
refine the decomposition of these two contributions by training an
ensemble of \gls{node} variants from different seeds.

% ============================================================
\section{Conclusion}
\label{sec:conclusion}
% ============================================================

We have presented a condition-monitoring framework for \gls{sic}
power modules that combines physics-informed cumulative features,
a Student-$t$ output head, and an optional monotonicity prior on
top of three compact neural architectures of increasing temporal
context. Evaluated under \gls{aqg}~$324$-compliant power cycling
on four operating-condition groups, the framework reduces the
mean absolute error on the lifetime-fraction estimate by up to
$60\%$ relative to a parameter-free baseline derived from the
qualification threshold, with the recurrent variant offering both
the best central tendency and the lowest cross-fold variance. The
dominant lever in the comparison is the feature engineering step:
encoding the cumulative thermo-mechanical history of the module
is what closes the gap between the raw per-cycle measurement and
a usable health estimate, and we read the result of this study as
a quantitative argument in favour of placing physical priors at
the input level rather than absorbing them entirely into
architectural complexity. A companion
study~\shortcite{scarpa2026iecon} benchmarks the same cumulative
\gls{node} against five reference estimators from the prognostics
literature: a parametric fit, a \gls{cnn}, a \gls{lstm}, a snapshot
\gls{mlp} and a transformer forecaster. The five are tested on two
campaigns with different failure mechanisms, the wire lift-off
studied here and solder degradation. The cumulative \gls{node} is
the only method whose accuracy transfers across both mechanisms,
while the reference estimators degrade as soon as the mechanism
changes. This confirms the weight of the input representation, and
of the junction-temperature history in particular, in making a
health estimate portable across failure modes. Broader operating envelopes that mix
mission profiles within a single trajectory, probabilistic metrics
that complement the point-estimate ones, and an online adaptation
phase that refreshes the network during normal converter
operation are the natural next steps to consolidate the framework
toward in-service deployment.

% ============================================================
%\section*{Acknowledgment}
% ============================================================
\section*{Acknowledgment}

In Germany this work of Infineon Technologies AG and Infineon Dresden AG \& Co KG is funded in the frame of the Important Project of Common European Interest on Microelectronics and Communication Technologies (IPCEI ME/CT). The IPCEI ME/CT is funded by the German Federal Ministry for Economic Affairs and Energy, the Bavarian Ministry for Economic Affairs, Regional Development and Energy, the Ministry of Economic Affairs, Industry, Climate Action and Energy of the State of North Rhine-Westphalia, the Saxon State Ministry for Economic Affairs, Labour, Energy and Climate Action and the European Union within “NextGenerationEU”.

% ============================================================
% Bibliography
% ============================================================
\bibliographystyle{apacite}
\bibliography{phm_sic_cm}
\nocite{*}

\end{document}